\documentclass[%
 aip,
 amsmath,amssymb,
 reprint,%
]{revtex4-1}

\usepackage{graphicx}
\usepackage{dcolumn}
\usepackage{bm}
\usepackage{hyperref}
\usepackage[utf8]{inputenc}
\usepackage[T1]{fontenc}
\usepackage{mathptmx}
\usepackage{etoolbox}

\makeatletter
\def\@email#1#2{%
 \endgroup
 \patchcmd{\titleblock@produce}
  {\frontmatter@RRAPformat}
  {\frontmatter@RRAPformat{\produce@RRAP{*#1\href{mailto:#2}{#2}}}\frontmatter@RRAPformat}
  {}{}
}%
\makeatother
\begin{document}

\title{A first-principles investigation of pressure induced topological phase transition in Half-Heusler AgSrBi}
\author{Bhautik R. Dhori}
\author{Raghottam M. Sattigeri}%
\author{Prafulla K. Jha}
\affiliation{Department of Physics, Faculty of Science, The Maharaja Sayajirao University of Baroda, Vadodara-390002, Gujarat, India}
\email{prafullaj@yahoo.com}
\author{Dominik Kurzydlowski}
\affiliation{Faculty of Mathematics and Natural Sciences, Cardinal Stefan Wyszyński University, ul. Wóycickiego 1/3, Warsaw 01-938, Poland.}%


\begin{abstract}
Topological Insulators (TI) are materials with novel quantum states which exhibit a bulk insulating gap while the edge/surface is conducting. This has been extensively explored in several Half-Heusler (HH) compounds hosting the exotic TI behaviour. In the present work we employ, first-principles based Density Functional Theory to perform thorough investigations of pressure induced topological phase transition (TPT) in HH AgSrBi which belongs to the F-43m space group. AgSrBi is intrinsically semi-metallic in nature which, under isotropic pressure exhibits semi-metal to trivial insulator transition retaining the cubic symmetry whereas, on breaking the cubic symmetry we observe the much sought after non-trivial semi-metal to TI phase transition. We also explore the effect of lowering crystal symmetry in realizing TI nature. Following this we perform qualitative analysis of the electronic properties to understand the origin of this non-trivial behavior followed by the quantitative analysis of the $\mathbb{Z}_2$ classification which indicates that AgSrBi is a strong TI (i.e., $\mathbb{Z}_2$ = (1, 101)). We thus, propose AgSrBi as a dynamically stable TI which can be used as ultra-thin films thermoelectric and nanoelectronic applications. 
\end{abstract}

\maketitle


The discovery of quantum spin hall effect (QSHE) \cite{1} and the governing topological properties of a material have sparked a lot of interest in the exotic and non-trivial phases of matter such as Topological Insulators (TI), Dirac semi-metals, Nodal-line semimetals etc. \cite{2,3,4,5,6,7} Followed by the theoretical prediction, the experimental realization of such materials has opened up numerous avenues from applications point of view which is not just restricted to spintronics, quantum computation but also extended to thermoelectricity, superconductivity etc. \cite{8,9,10} TI are essentially insulating in D dimension and conducting in D-1 dimension i.e., in three dimensional regime, the bulk is insulating but the surfaces are conducting in nature. This is by the virtue of the time reversal symmetry which makes these conducting surface states topologically robust against external perturbations. Such behavior was initially observed in the bulk quintuple layered system Bi$_{1-x}$Sb$_x$, \cite{11,12} which was preceded by numerous other compounds with similar structure such as Bi$_2$Te$_3$,Sb$_2$Te$_3$ \cite{13} etc. However, the search for such non-trivial topological properties was further explored in other crystal structures such as, binary compounds of group IV-VI compounds, magnetic compounds such as AMgBi (A = Li, Na, K)\cite{14,15} and, Full and Half-Heusler (HH) compounds (which are also known for their thermoelectric, piezoelectric, magnetic and superconducting properties). \cite{16,17,18,19,20,21}

\begin{figure*}[ht]
	\centering
	\includegraphics[width = 16cm]{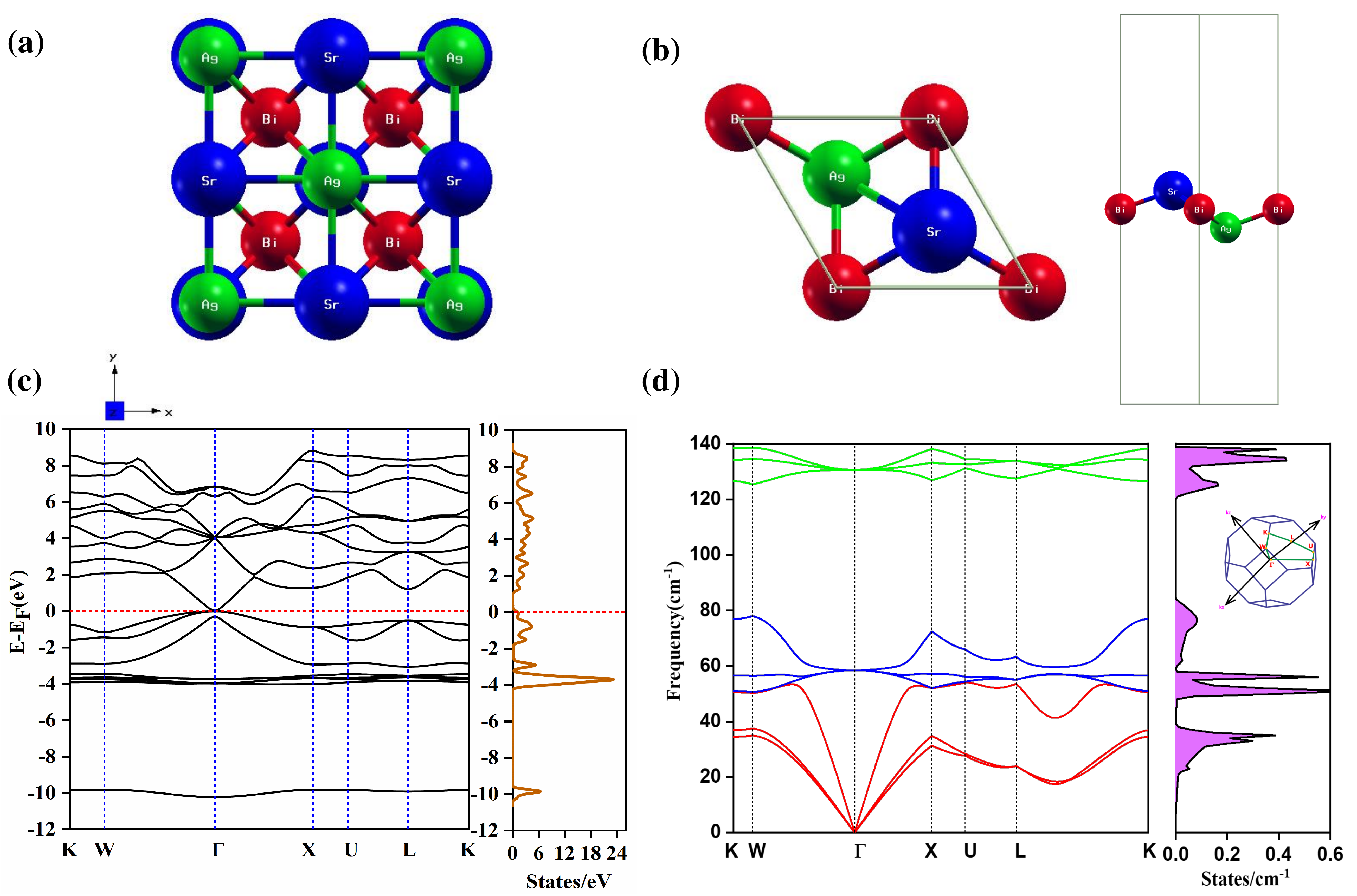}
	\caption{\label{Figure 1:} (a) Face centred cubic (fcc) structure of AgSrBi (b) [1 1 1] plane of FCC AgSrBi (c) electronic band structure without spin orbit coupling alongside density of state (DOS) indicating semi-metallic nature of AgSrBi (d) phonon dispersion curve alongside of phonon density of state indicating absence of imaginary modes in the entire brillouin zone which implies of AgSrBi is dynamically stable}
\end{figure*}
 
The exciting possibility of multifunctional properties alongside the TI property makes HH materials ideal candidates from applications point of view. HH compounds also exist is different polytypes such as hexagonal, orthorhombic, cubic etc. \cite{22} However, the cubic structure belonging to F$\overline{4}$3m space group is widely explored from topological perspective since, the zincblende sub-lattice gives rise to band orders which are similar to that of HgTe since they are connected via topological adiabatic connection. \cite{23} Also, the topological character in a typical HH ABC crystal arises due to the cation (A) substitution in a anion (BC) zincblende sub-lattice. Hence, HH compounds are potential candidates which can host rich and diverse properties of interest \cite{16,17,18,19,20,21}. However, not every HH compound exhibits the TI nature intrinsically as observed in Bi$_2$Te$_3$ and Sb$_2$Te$_3$ \cite{13} which arises due to the strong influence of spin-orbit coupling (SOC) between the core and valence electrons giving rise to relativistic effects, rather, HH compounds exhibit TI character under different regimes such as, influence of pressure, electric field, doping, vacancy and dimensional engineering. \cite{24,25,26,27,28,29} These mechanisms have been explored in various HH compounds, however, HH based on silver are quite unexplored theoretically as well as experimentally leading to scarcity of data. 

HH systems such as, AgSrX with X = As, Sb, Bi have been predicted to be thermodynamically stable \cite{22}, however, there is a lack of clarity on their electronic and topological properties. With AgSrAs \cite{17} being an exception in this regard, since it has been explored for its, structural, piezoelectric properties. AgSrBi unravels some unconventional electronic property with semi-metallic behavior. This inspired us to explore it in terms of the SOC interactions and its influence on the topological character of AgSrBi. Thus, in the present study we explore the structural, electronic and topological properties of AgSrBi. We observe that, the semi-metallic character of AgSrBi is retained even when SOC is considered. This motivated us to explore different potential routes to realise the non-trivial topological nature in AgSrBi. This is achieved by, (i) breaking the cubic symmetry and (ii) lowering the crystal symmetry along with dimensional confinement along [001] crystal direction (which gives rise to the quantum confinement effects) \cite{30,31,32}. The former mechanism is known to lift off the degeneracies at the Fermi while the later gives rise to quantum spin hall effects which host the robust surface and edge states respectively. With a focus on the topological character of AgSrBi, we perform qualitative analysis of the electronic band structure followed by the quantitative analysis of the $\mathbb{Z}_2$  topological class. We find that, AgSrBi can host non-trivial TI nature when, the cubic symmetry is broken and crystal symmetry is lowered to hexagonal phase from face centered cubic phase along with dimensional confinement along [001] crystal direction. We further classify this with $\mathbb{Z}_2$  invariant as (0, 1 2 3) = (1, 101) and  = 1 (and non-zero Chern number) which indicates the strong and non-trivial TI nature respectively. This suggests that, the thin films of AgSrBi can be either exfoliated or synthesized by using molecular beam epitaxy for applications in nanoelectronics and spintronics. \cite{33,34}


We use first-principles based Density Functional Theory (DFT) implemented in Quantum ESPRESSO code \cite{35} to calculate and investigate the electronic properties of AgSrBi. We use norm conserving pseudopotentials under generalized gradient approximation based on Martins-Troullier method \cite{36} with exchange correlation energy functional of Perdew-Burke-Ernzerhof type.\cite{37} This pseudopotential considers, 4d$^10$5s$_1$,5s$^2$ and 6s$^2$6p$^3$ orbitals of As, Sr and Bi respectively. In ordered to consider the relativistic effect from core electrons on the valence electrons in SOC calculations, pseudopotential based on projector augmented wave (PAW) method was used. The ground state of the system was optimized by following the bisection convergence method to locate the global minima with convergence threshold of the order of < 10$^-8$ Ry. The kinetic energy cutoff of 80 Ry was adopted for the basis set with the irreducible Brillouin Zone (BZ) sampled on a uniform Monkhorst-pack grid \cite{38} k-mesh of 8 $\times$ 8 $\times$ 8 and 8 $\times$ 8 $\times$ 1. The vibrational properties were calculated using the Density Functional Perturbation Theory (DFPT) \cite{39} with a q-mesh of 6 $\times$ 6 $\times$ 6. Finally, we generate a tight binding model for the system by using the maximally localized wannier functions which were obtained using Wannier90 \cite{40,41} and then parsed to the WannierTools \cite{42} wherein the $\mathbb{Z}_2$ classification (and Chern number calculations) was performed around the wannier charge centers followed by plotting the surface and edge states using the iterative Green’s function method.


\begin{figure*}
	\centering
	\includegraphics[width= 16cm]{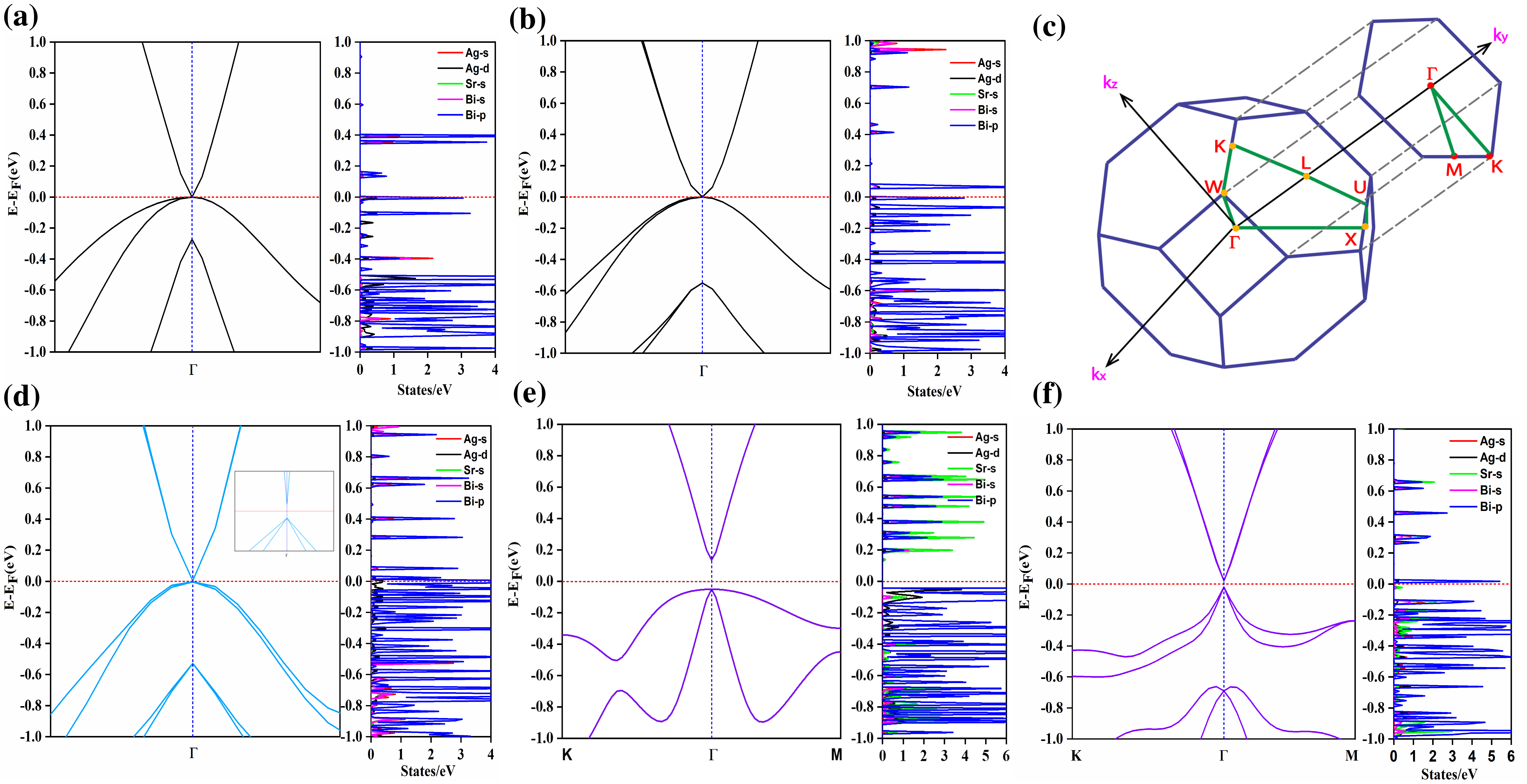}
	\caption{\label{Figure 2:} Electronic band structure of AgSrBi along side of partial density of state (PDOS) without SOC (a) with SOC (b). (c) Irreducible first Brillouin Zones with the path for Electronic Band Structure indicated for the bulk and low-dimensional phase of AgSrBi.  (d) electronic band structure of AgSrBi (inset, non-trivial gap) with broken symmetry due to uniaxial strain alongside PDOS indicating non-trivial orbital inversion. Electronic Band Structure of the low-dimensional [1 1 1] phase of AgSrBi alongside of Partial Density of States (PDOS) (e) Without SOC (f) With SOC.}
\end{figure*}


We begin with the description of the bulk HH AgSrBi which is predicted to exhibit a face-centered cubic structure governed by the F$\overline{4}$3m [216] space group (as shown in Fig. \ref{Figure 1:}(a)) with the primitive cell vectors defined in terms of Wyckoff positions of Ag, Sr and Bi as, v1 =  (a/2)(-1,0,1) , v2 = (a/2)(0,1,1)  and v3 = (a/2)(-1,1,0) respectively. Based on convergence tests, the optimized lattice constant of a = 7.306 \AA indicates the global minima regime. This phase of AgSrBi exhibits semi-metallic nature which is unusual as compared to AgSrAs. This structure is found to be thermally stable since the formation energy is – 0.52 eV which implies that the system will form spontaneously when appropriate precursors are combined.

Since there is lack of experimental and theoretical data on the vibrational modes of AgSrBi, we proceed with the investigation of phonon modes which would give insights into the lattice dynamics. It is evident from Fig. \ref{Figure 1:}(d) that, AgSrBi does not exhibit any negative/imaginary frequency in the BZ (presented in Fig. \ref{Figure 2:}(c)) indicating dynamical stability of AgSrBi under pristine conditions in the harmonic approximation regime. There are a total of nine phonon branches with three acoustic and six branches optical phonons due to the three atoms in the primitive cell. The lower frequency regime is governed by heavier Ag and Bi atoms whereas the higher frequency regime is governed by the lighter Sr atom with slight contributions from the Ag atom. At zero pressure which corresponds to the pristine conditions, the acoustic branches lie around 50-60 $cm^{-1}$, this implies that, AgSrBi would exhibit isotropically low thermal conductivity ($\kappa$) which is a key factor from thermoelectric perspective (since,  depends quadratically on the slope of the acoustic branches). Also, since all the branches at zero pressure lie below 200 $cm^{-1}$, we can also expect the system to be stable at higher temperatures owing to the vibrational entropy. Apart from this, a peculiar feature at zero pressure is the degeneracy and mixing of the acoustic and optical phonon branches which indicates the semi-metallic behaviour of the material and the similarity of the system with rock-salt oxides. The degenerate longitudinal optical (LO) and transverse optical (TO) branches at the zone center $\Gamma$ indicate the absence of long range dipole-dipole interaction and hence confirm the non-semiconducting character of the system. We scan the phonon dispersion curves by varying the pressure from 0.0 GPa to 25.0 GPa and find that the system is dynamically stable throughout. However, there is an increment in the softening between L - K points in the BZ which implies that, higher pressures would eventually make the system dynamically unstable.

On lowering the crystal symmetry from face centered cubic to hexagonal symmetry along the [111] crystal direction of the bulk and confining the system along [001] crystal direction, we obtain a crystal structure similar to 1T-$MoS_{2}$. This low dimensional phase of AgSrBi has optimized lattice constant of a = 4.28 \AA (Fig. \ref{Figure 1:}(b)). The atomic arrangement is of sandwich type similar to 1T-$MoS_{2}$ with Ag and Sr atoms occupying the top and bottom layers and the Bi atoms occupying the middle layer as presented in Fig. \ref{Figure 1:}(b). The height of this sandwich is 2.11 \AA with interatomic distances as, 3.65 \AA, 3.16 \AA and 3.16 \AA for Ag-Sr, Ag-Bi and Sr-Bi respectively. The confinement is achieved by introducing a vacuum of 15\AA along [001] crystal direction. The cohesive energy of this system is - 6.94 eV/atom; which is comparable to graphene \cite{43}, indicating exfoliation from the [111] crystal plane as a potential route to experimentally realise the system other than molecular beam epitaxy.

\begin{figure*}[ht]
		\centering
		\includegraphics[width = 15cm]{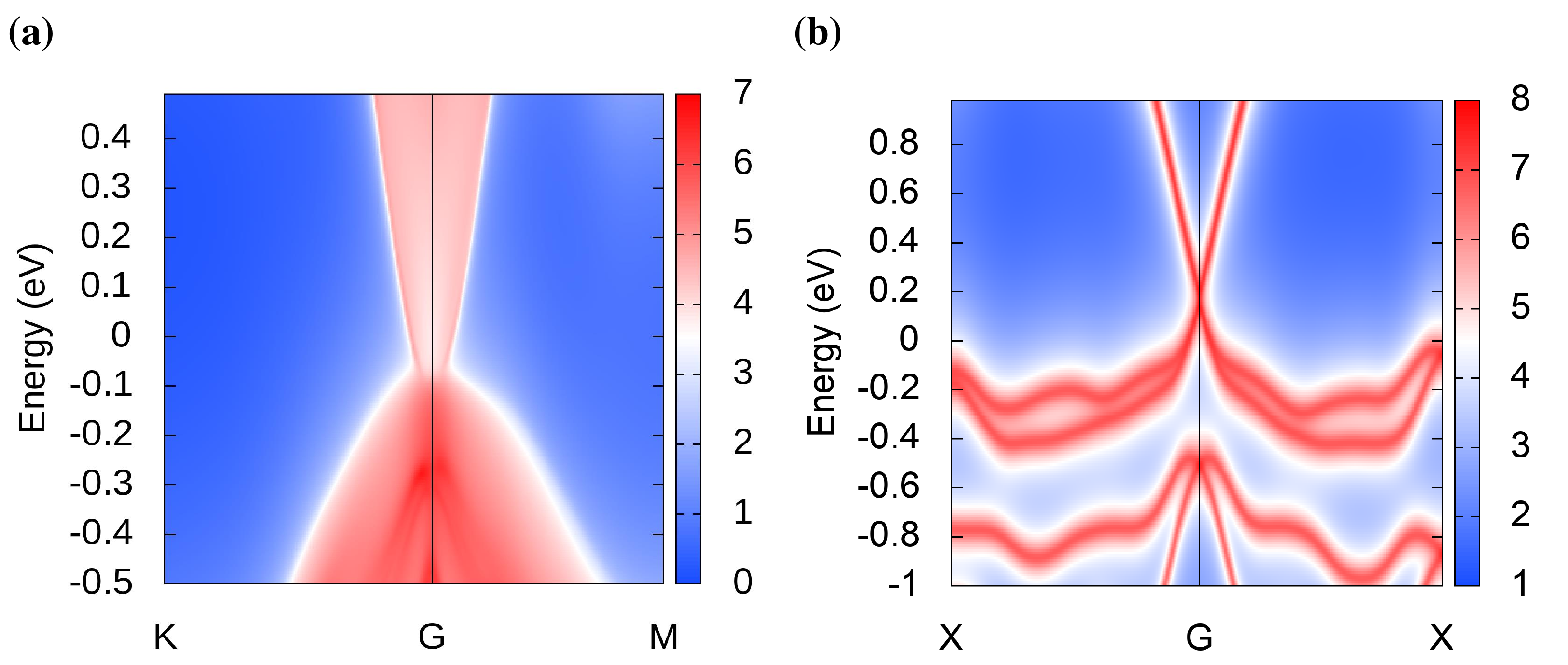}
		\caption{\label{Figure 3:}Computational Surface and edge states of the (a) Bulk and (b) Low dimensional phase of HH AgSrBi respectively (heat map of the spectrum is in arbitrary unit).}
\end{figure*}


Under pristine conditions i.e., at zero pressure in the absence of SOC, bulk AgSrBi exhibits a semi-metallic band structure with degenerate valence and conduction bands crossing each other at Fermi along the high symmetry point $\Gamma$ in the entire BZ (as evident from Fig. \ref{Figure 1:}(c)). We restrict our investigation of the electronic band structure in the vicinity of Fermi level since, the low energy electronic properties of materials can be determined by the nature of bands located in this region. From Fig. \ref{Figure 2:}(a), it is evident that, the valence band maxima has a doubly degenerate state along the high symmetry point $\Gamma$, whereas, the conduction band minima has single non-degenerate state which is protected by the two-fold symmetry of the face-centered cubic structure. From the partial density of states it is clear that the valence band is populated by the -s, -p, -d orbitals of Ag, -s orbitals of Sr and -s, -p orbitals of Bi; whereas, the conduction band is populated by, -s orbitals of Ag, -s orbitals of Sr and -s, -p orbitals of Bi.  When SOC is imposed, the band degeneracies are not altered; retaining their semi-metallic character (evident from Fig. \ref{Figure 2:}(b)). However, when an external uniaxial compressive pressure is applied to the system, the cubic symmetry is broken and the orbital character is altered. This enhances the SOC effect leading to the non-trivial TI phase. Breaking the cubic symmetry of a system has been one of the effective methods to realize a non-trivial gap (in the otherwise degenerate states) leading to the much sort after TI nature \cite{44,45}. When we subjected to a uniaxial compressive strain (in steps of 0.1\%) along [100] crystal direction (in the presence of SOC) we observe non-trivial behavior. At 0.2\% (1.5 GPa) strain itself the degeneracy at $\Gamma$ lifts off but, the conduction band encapsulates the Fermi level indicating the Topological Conductor behavior. However, on further increasing the strain we arrive at a critical strain of 1.3\% (1.9 GPa) which gives rise to a non-trivial gap of 0.35 meV with Fermi level lying in between the conduction and valence bands (presented in Fig. \ref{Figure 2:}(d)). This gap is global in nature and retained throughout the BZ indicating towards the possibility of low-dimensional conducting surface states. The band inversion mechanism behind this non-trivial gap is due to the new contributions in conduction band arising due to the -s orbitals of Ag, Sr and Bi and -p orbitals of Bi while the contributions from -d orbitals of Ag are absent and do not contribute. This makes it clear that, the band inversion is of -s-p type which is quite common in several HH compounds and their characteristic behaviour indicating a non-trivial TI phase. 

Another route to realize non-trivial topological behavior in materials is by lowering the crystal symmetry and confining the system in a particular crystal direction. This method has been previously explored in HgTe binary compound \cite{46}. The bulk phase (with cubic symmetry) of HgTe does not exhibit the topological character, whereas, on lowering the crystal symmetry (to a hexagonal symmetry) and confining the system in one dimension gave rise to interesting topological properties. The lower crystal symmetry was achieved by cleaving [111] plane of bulk HgTe compound and confining it along the [001] crystal direction. This opened the gap in the degenerate electronic structure (as compared to bulk) but, the band ordering was trivial. This was addressed by designing quantum wells of CdTe and HgTe with the thickness of quantum well governing the critical regime of the system which would host non-trivial topology characterized by inverted band order.

In case of AgSrBi, we expect the nature of electronic structure to be analogous to low dimensional HgTe which as compared to its bulk phase exhibits TI nature. For this purpose, we reduce the crystal symmetry of bulk AgSrBi and confine the resulting system into one dimension along [001] crystal direction. Similar to HgTe, the low dimensional phase of AgSrBi exhibits a semi-conducting nature with a global gap of 0.19 eV along the high symmetry point $\Gamma$ in the BZ (as evident from Fig. \ref{Figure 2:}(e)). In the absence of external stress on the system, we impose SOC effects into our calculations and find that the gap reduces to 39 meV along the high symmetry point $\Gamma$ in the BZ (as evident from Fig. \ref{Figure 2:}(f)). This gap is non-trivial in nature and superior to some previously reported systems \cite{47,48}. However, this gap can be further enhanced by subjecting the system to strain, functionalization (partial/complete) and by creating multilayer system \cite{49,50}. The non-trivial nature of the gap is clearly evident from the projected density of states presented in Fig. \ref{Figure 2:}(f). In the absence of SOC, the valence band maxima is dominated by -p orbitals of Bi atom with minor contributions from -d orbitals of Ag and -s orbitals of Sr and Bi, whereas the conduction band minima is dominated by the -s orbital of Sr and minor contributions from the -p orbitals of Bi (as evident from Fig. \ref{Figure 2:}(e)). On imposing SOC we observe a clear band inversion as the -p orbitals of Bi shift into the conduction band minima giving rise to the non-trivial nature. We attribute this phenomena to the quantum confinement effects and the extra degrees of freedom to electrons from the -p orbitals arising from the vacuum along [001] direction. 

We now classify the proposed system into $\mathbb{Z}_2$ topological class followed by computing the surface and edge states using the iterative Green’s function method.


The non-trivial topology of materials is classified in terms of the $\mathbb{Z}_2$ invariant. Typically, the $\mathbb{Z}_2$ invariants for materials with inversion symmetry are calculated as the product of parities of eigen values at Time Reversal Invariant Momenta (TRIM) points. However, since HH compounds like AgSrBi do not possess inversion symmetry (owing to the cubic crystal structure), the $\mathbb{Z}_2$ invariant is computed along two different momentum planes in the BZ as $\left(\mathbb{Z}_{2}\right)_{\left(k =\pi\right)}$ and $\left(\mathbb{Z}_{2}\right)_{\left(k =0\right)}$. The $\mathbb{Z}_2$ invariants are calculated in terms of the wannier charge center in the vicinity of Fermi level using WannierTools code which utilizes the wannier functions encoded in the tight binding model obtained from the Wannier90 code. The $\mathbb{Z}_2$ invariants are calculated along six Time Reversal Invariant Planes (TRIP) i.e.,k$_x$ = 0,$\pi$, k$_y$ = 0,$\pi$ and k$_z$ = 0,$\pi$ in the BZ. At a critical uniaxial strain of 1.3\% (1.9 GPa) which corresponds to the non-trivial gap (presented in Fig. \ref{Figure 2:}(d)), we compute the $\mathbb{Z}_2$ invariants as, $\nu_0 = \Big(\mathbb{Z}_2 \big(k_i = 0 \big) + \mathbb{Z}_2 \big(k_i = 0.5 \big) \Big) mod 2$ and $\nu_i = \mathbb{Z}_2 \big(k_i = 0.5 \big)$ and find them to be ($\nu_0$, $\nu_1$ $\nu_2$ $\nu_3$) = (1, 1 0 1). This indicates strong topological insulating nature of bulk AgSrBi. The corresponding surface states are computed using iterative Green’s function method along [111] crystal direction and shown in Fig. \ref{Figure 3:}(a). This makes it clear that, the bulk gap in AgSrBi at critical uniaxial strain hosts the non-trivial surface states which are robust against external perturbations. Similarly, the non-trivial nature of the lower crystal symmetry and dimensionally confined system is characterized by the $\mathbb{Z}_2$ invariant as $\nu$ = 1 with a non-zero Chern number ($\mathcal{C}$ = 1) indicating the potential applications as quantum spin Hall insulators. From the edge states computated along the orthorhombic edge BZ presented in Fig. \ref{Figure 3:}(b), it is clearly evident that, the global gap along high symmetry point $\Gamma$ in the BZ exhibits non-trivial quantum spin hall insulating nature.


To sum up, we employ first-principles based DFT calculations to investigate TI character of dynamically stable HH compound AgSrBi. Under pristine conditions (zero pressure) AgSrBi exhibits semimetallic behavior which is retained even when SOC is considered. However, when subjected to isotropic pressure the system retains this behavior but, in order to realize non-trivial nature, we break the cubic crystal symmetry and lower the symmetry along with dimensional confinement giving rise to the TI nature. This TI nature hosts conducting surface states along [111] direction evident from the surface states while the bulk hosts a global gap. Also, this non-trivial low dimensional phase is classified to be a strong TI with $\mathbb{Z}_2$ invariant as (1, 1 0 1). Similarly, on lowering the crystal symmetry analogous to HgTe; we observe a non-trivial TI nature characterized by the band inversion, $\mathbb{Z}_2$ invariant ($\nu$ = 1), non-zero Chern number ($\mathcal{C}$ = 1) and the edge states. Our results propose yet another potential TI and quantum spin Hall insulator with spintronic degrees of freedom to the electrons and potential nanoelectronic applications.

\section*{ACKNOWLEDGEMENT}
This research was carried out with the support of the Interdisciplinary Centre for Mathematical and Computational Modelling (ICM), University of Warsaw, Poland, under Grants no. GB84-24 and GA83-26.

\section*{References}
\nocite{*}
\bibliography{aipsamp}

\end{document}